\documentclass[sn-mathphys,Numbered]{sn-jnl}

\usepackage{graphicx}%
\usepackage{multirow}%
\usepackage{amsmath,amssymb,amsfonts}%
\usepackage{amsthm}%
\usepackage{mathrsfs}%
\usepackage[title]{appendix}%
\usepackage{xcolor}%
\usepackage{textcomp}%
\usepackage{manyfoot}%
\usepackage{booktabs}%
\usepackage{algorithm}%
\usepackage{algorithmicx}%
\usepackage{algpseudocode}%
\usepackage{listings}%
\usepackage{bm}
\usepackage{siunitx}
\usepackage{framed}
\usepackage{braket}
\usepackage{soul}

\theoremstyle{thmstyleone}%
\theoremstyle{thmstyletwo}%

\theoremstyle{thmstylethree}%

\begin{document}

\title[Article Title]{Quantum Geometric Renormalization of the Hall Coefficient and Unconventional Hall Resistivity in
$\mathrm{ZrTe}_{5}$}


\author[1,2]{\fnm{Xie} \sur{Huimin}}

\author*[2]{\fnm{Fu} \sur{Bo}}\email{fubo@gbu.edu.cn}

\author*[3]{\fnm{Wang} \sur{Huan-Wen}}\email{wanghw@uestc.edu.cn}

\author*[1]{\fnm{Shan} \sur{Wenyu}}\email{wyshan@gzhu.edu.cn}

\author[4]{\fnm{Shen} \sur{Shun-Qing}}

\affil[1]{\orgdiv{Department of Physics,School of Physics and Materials Science}, \orgname{Guangzhou
University}, \orgaddress{\city{Guangzhou}, \postcode{510000}, \state{Guangdong Province}, \country{China}}}

\affil[2]{\orgdiv{School of Sciences}, \orgname{Great Bay University}, \orgaddress{\city{Dongguan}, \postcode{523000}, \state{Guangdong Province}, \country{China}}}

\affil[3]{\orgdiv{School of Physics}, \orgname{ University of Electronic Science and Technology
of China}, \orgaddress{\city{Chengdu}, \postcode{610000}, \state{Sichuang Province}, \country{China}}}

\affil[4]{\orgdiv{Department of Physics}, \orgname{The University of Hong Kong}, \orgaddress{\street{Pokfulam Road}, \city{Hong Kong},   \country{China}}}



\abstract{The anomalous Hall effect (AHE), conventionally associated with time-reversal
symmetry breaking in ferromagnetic materials, has recently been observed
in nonmagnetic topological materials, raising questions about its
origin. We unravel the unconventional Hall response in the nonmagnetic
Dirac material $\mathrm{ZrTe}_{5}$, known for its massive Dirac bands
and unique electronic and transport properties. Using the Kubo-Streda
formula within the Landau level framework, we explore the interplay
of quantum effects induced by the magnetic field ($B$) and disorder
across the semiclassical and quantum regimes. In the semiclassical
regime, the Hall resistivity remains linear in the magnetic field,
but the Hall coefficient will be renormalized by the quantum geometric
effects and electron-hole coherence, especially at low carrier densities
where the disorder scattering dominates. In quantum limit, the Hall
conductivity exhibits an unsaturating 1/B scaling. As a result, the
transverse conductivity dominates transport in the ultra-quantum limit,
and the Hall resistivity crosses over from $B$ to $B^{-1}$ dependence
as the system transitions from the semiclassical regime to the quantum
limit. This work elucidates the mechanisms underlying the unconventional
Hall effect in $\mathrm{ZrTe}_{5}$ and provides insights into the
AHE in other nonmagnetic Dirac materials as well.}

\maketitle

\section{Introduction}\label{sec1}

The anomalous Hall effect (AHE) is a key electrical transport phenomenon
with significant implications for both fundamental physics and applications
\textsuperscript{\citep{haldane1988model,nagaosa2010anomalous,chang2013experimental,burkov2014anomalous,checkelsky2014trajectory,burkov2018weyl,deng2020quantum,Bernevig2022Progress,zhang2024nonmonotonic,belopolski2025synthesis}}.
First observed in ferromagnetic iron\textsuperscript{\citep{hall1881possibility}},
the microscopic mechanisms of AHE have been debated for nearly a century\textsuperscript{\citep{karplus1954hall,sinitsyn2007anomalous,onoda2006intrinsic,xiao2010berry}}.
Typically, AHE requires time-reversal symmetry breaking via magnetism
with Hall resistivity as $\rho_{\mathrm{xy}}=R_{0}B+R_{\mathrm{AH}}M$, where $R_{0}B$
represents  the magnetic field ($B$) linear ordinary Hall effect
and $R_\mathrm{AH}M$ corresponds to the magnetization induced anomalous
Hall effect.

$\mathrm{ZrTe}_{5}$ is a nonmagnetic topological material characterized
by massive Dirac bands, situated at the boundary between strong and
weak topological insulators\textsuperscript{\citep{weng2014transition,zhang2019anomalous,jiang2020unraveling,chen2015magnetoinfrared,li2016chiral,mutch2019evidence,zhang2021observation,tang2019three}}.
It exhibits a paramagnetic response at low magnetic fields and no
signatures of magnetic interactions \textsuperscript{\citep{nair2018thermodynamic,ji2021berry}}.
A variety of intriguing phenomena have been observed in this material,
including log-periodic quantum oscillations\textsuperscript{\citep{wang2018discovery}},
3D quantum Hall effects\textsuperscript{\citep{tang2019three,qin2020theory,galeski2021origin}},
resistivity anomaly\textsuperscript{\citep{okada1980giant,izumi1981anomalous,tritt1999large,rubinstein1999hfte,shahi2018bipolar,wang2021thermodynamically,fu2020dirac}},
and negative magnetoresistance\textsuperscript{\citep{li2016chiral,wang2021helical}}.
Recently, an unconventional Hall signal has been reported in $\mathrm{ZrTe}_{5}$:
the Hall resistivity $\rho_\mathrm{xy}$ exhibits an unconventional behavior in
the high-field regime\textsuperscript{\citep{liang2018anomalous,gourgout2022magnetic,choi2020zeeman,mutch2021abrupt,liu2021induced,sun2020large,wang2023theory,wang2023quantum,pi2024first,wu2023topological,2022Signatures,bj2n-4k2w}}.
This behavior, reminiscent of the anomalous Hall effect, is frequently
attributed to the Berry curvature of the electronic bands, potentially
arising from Zeeman splitting or the formation of Weyl nodes \textsuperscript{\citep{sun2020large,zhao2023magnetotransport,liu2021induced}}.
However, most studies rely on a semiclassical approximation and often
neglect the orbital effects of the magnetic field. This oversight
is particularly significant in systems with narrow band gap and low
carrier density, where the influence of the orbital effect of  the magnetic
field and disorder scattering becomes pronounced. The origin of the
unconventional Hall effect in paramagnetic topological materials like
$\mathrm{ZrTe}_{5}$ remains under debate, necessitating a quantitative
investigation to clarify underlying mechanisms.

\begin{figure}
\includegraphics[width=12cm]{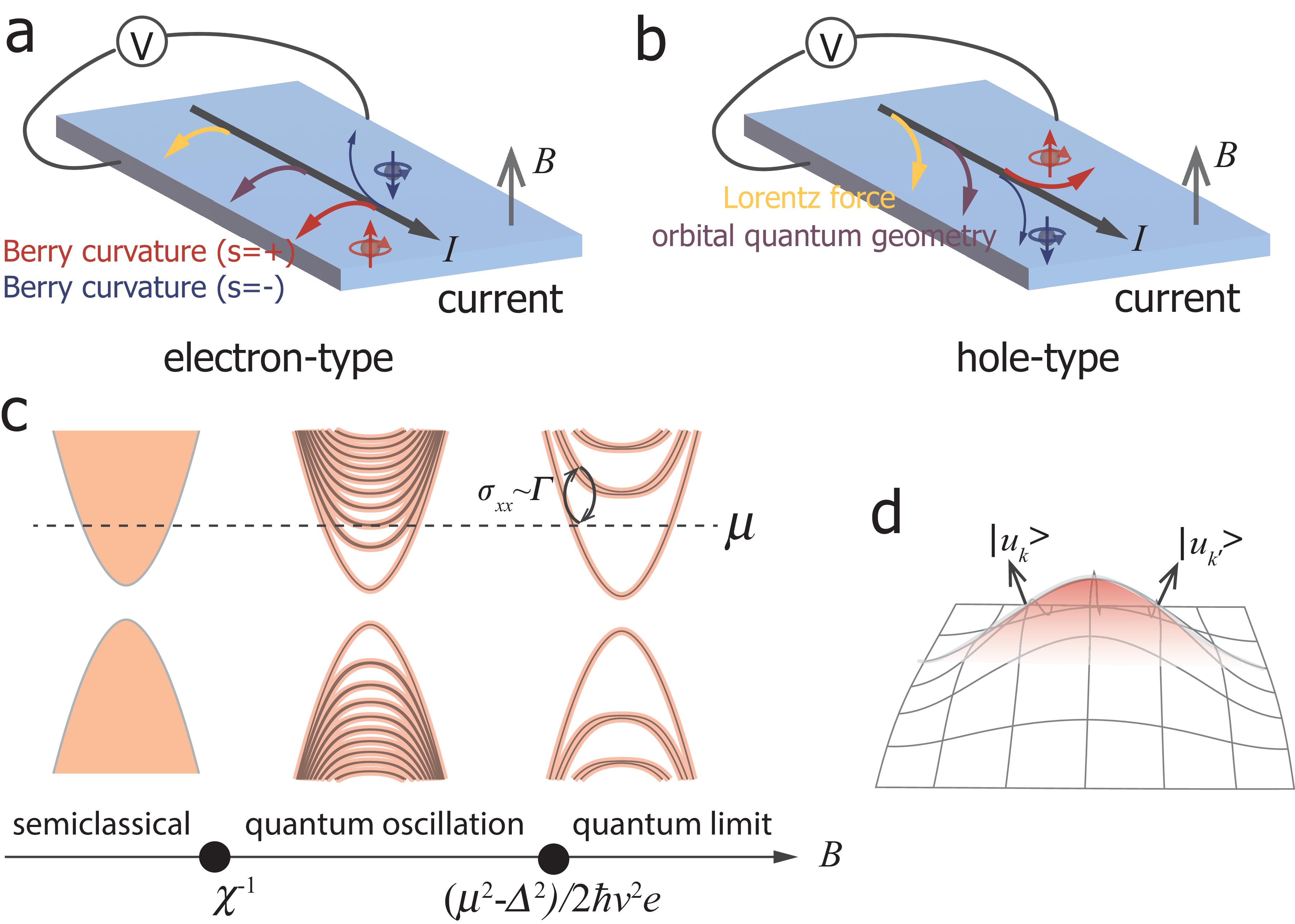}

\caption{\protect\label{fig:illustration_diagram}\textbf{Schematic illustration of Hall conductivity mechanisms in Dirac materials and three magnetic field regimes.}The Hall effect in Dirac materials arises from distinct contributions for \textbf{a} electron-type and \textbf{b} hole-type carries, where a longitudinal current $I$ under a perpendicular magnetic field $B$ generates a transverse voltage $V$. Arrows indicate carrier motion directions: yellow for the classical Lorentz force and brown for quantum geometric effects (e.g., quantum metric and orbital magnetization). Under a fixed $I$, the Lorentz and quantum geometric terms produce transverse velocities that are invariant under carrier sign reversal, leading to an inverted Hall voltage.   Red and blue arrows denote the Berry curvature contributions from Zeeman-split majority ($s=+$) and minority $s=-$ states, respectively. The Berry curvature induces opposite anomalous velocities for $s=\pm$, creating a transverse carrier imbalance and a finite voltage. Unlike the classical and geometric terms, this contribution reverses under carrier sign change, leaving the Hall voltage unchanged. \textbf{c} Evolution of electronic states with magnetic field. The system progresses through three regimes: (i) semiclassical ($B<\chi^{-1}$)(ii) quantum oscillations, and (iii) quantum limit  ($B>(\mu^2-\Delta^2)/(2\hbar v^2e)$). In the semiclassical regime, disorder broadening smears out the Landau levels. In the quantum limit, all carriers occupy only the lowest Landau level. Disorder broadening (red shading) and chemical potential $\mu$ (black dashed line) are shown for reference. \textbf{d} Quantum geometric effect-induced Hall effect. The black arrows depict the Bloch wavefunctions at adjacent $k$-points, with their directional difference representing the quantum metric (state distance). The red shading indicates the emergent curvature from interband coupling.}
\end{figure}

In this work, we investigate the mechanism of the unconventional Hall
effect in paramagnetic Dirac materials by employing the Kubo-Streda
formula within the framework of Landau levels, which deals  with the quantum
effect of magnetic fields and disorder on an equal footing. As shown in Fig. \ref{fig:illustration_diagram}\textbf{a} and \textbf{b}, our calculations in the semiclassical regime reveal that quantum geometry effects—such as the Berry curvature (red/blue arrows) and quantum metric contributions (brown arrows)—introduce significant quantum corrections to the classical Lorentz-force-driven Hall conductivity (yellow arrows). 
These effects renormalize the Hall coefficient, particularly at low
carrier densities where disorder scattering plays a significant role. Owing to their distinct symmetry properties under particle-hole transformation, the quantum geometry effects exhibit characteristically different behavior for electron- and hole-type carriers.
In the quantum oscillation regime, the Fermi-surface contribution of the Zeeman-induced Berry curvature is suppressed by magnetic orbital effects and exhibits oscillations due to Landau level quantization, while the Fermi-sea contribution remains robust.
In the quantum limit where the quasiclassical picture is entirely invalid,
although the Zeeman splitting drives the formation of Weyl nodes,
the Hall conductivity scales as $\sim en/B$ without saturation, where
$n$ is the carrier density. Consequently, the transverse conductivity
dominates transport in the ultra-quantum limit, causing the Hall resistivity
to be inversely proportional to $B$. The crossover from the semiclassical
regime to the quantum limit results in unconventional Hall resistivity,
and we identify the critical magnetic field at which the Hall resistivity
transitions from positive to negative. This study presents a unified
framework to elucidate the intricate interplay of quantum geometry,
magnetic fields, and disorder in paramagnetic Dirac materials, while
also shedding light on the mechanisms behind anomalous transport phenomena
in numerous nonmagnetic Dirac systems.

\begin{figure}
\includegraphics[width=12cm]{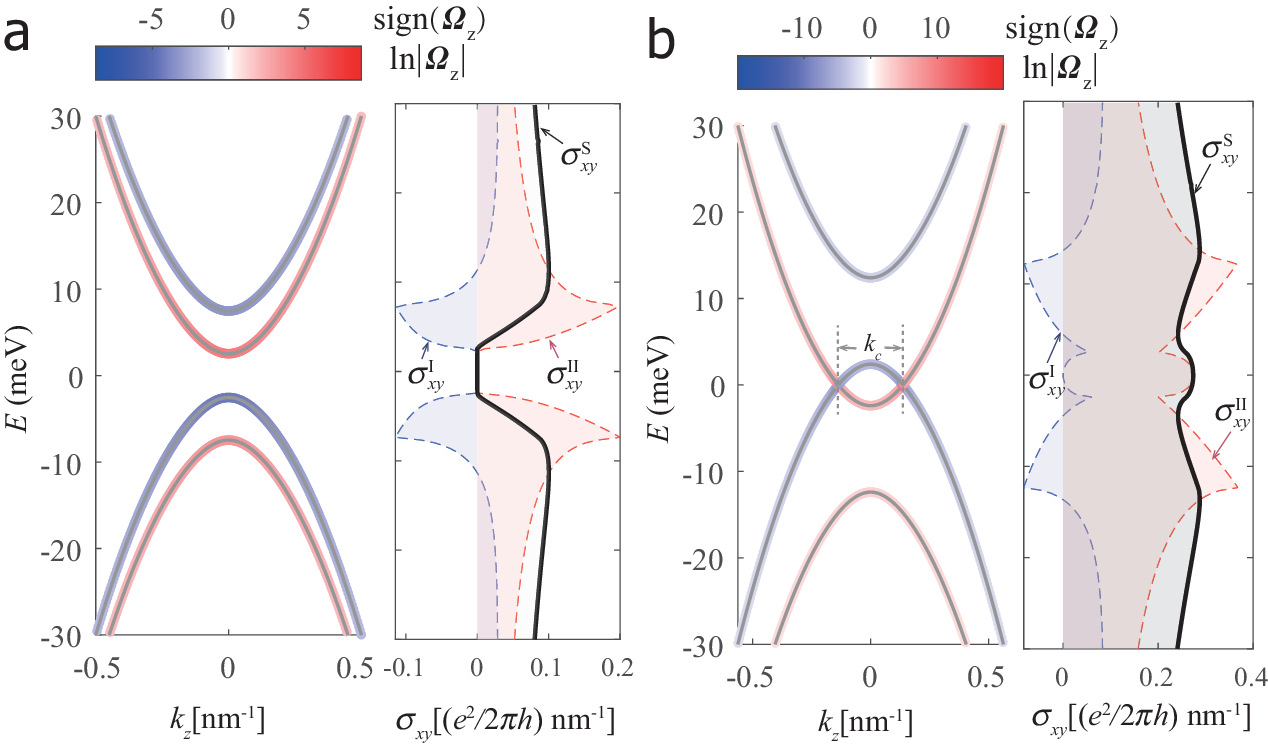}

\caption{\protect\label{fig:onlyZeeman}\textbf{Zeeman Splitting Modifications to Electronic Band Structure and Hall Transport} Band structure $\varepsilon_\mathrm{s\zeta}(\mathbf{k})$ at
$|\mathbf{k}_{\perp}|=0$ and  the Zeeman-splitting-induced Hall effect for two magnetic field strengths:  \textbf{a} $B=4$T and \textbf{b} $B=12$ T, below and above the  field $B_\Delta=2\Delta/(g_\mathrm{z}\mu_\mathrm{B})\approx8.11T$  required for Weyl nodes formation. The electronic structure is color-coded (red to blue) to represent the logarithmic Berry curvature, $\mathrm{sign}(\Omega_z)\ln|\Omega_\mathrm{z}|$. In \textbf{b}, $k_c$  denotes the momentum-space separation between the Weyl nodes. The right panels of each subfigure show the Hall conductivity components: $\sigma_\mathrm{xy}^{\mathrm{I}}$ (blue dashed), $\sigma_\mathrm{xy}^{\mathrm{II}}$ (red dashed), and their sum $\sigma_\mathrm{xy}^{\mathrm{S}}$(black solid) versus energy $E$. 
Model parameters: $v_\mathrm{x}=9.11\times10^{5}\mathrm{m\cdot s^{-1}}$,
$v_\mathrm{y}=1.97\times10^{4}\,\mathrm{m\cdot s^{-1}}$, $t_\mathrm{z}=20$ meV,
$\Delta=5$ meV, $C=100$ meV, $a=1$ nm, $g_\mathrm{z}=21.3$, $\mu_\mathrm{B}=5.788\times10^{-2}$
$\mathrm{meV}\cdot$$\mathrm{T}^{-1}$.}
\end{figure}
\section{Results}\label{sec2}

\paragraph*{Model and Zeeman Effects}

The anisotropic Hamiltonian for $\mathrm{Zr}\mathrm{Te}_{5}$ in a
finite perpendicular magnetic field $\mathbf{B}$, can be written
as \textsuperscript{\citep{chen2015magnetoinfrared}}
\begin{equation}
H(k)=\Delta(k_\mathrm{z})\tau_\mathrm{z}+m\sigma_\mathrm{z}+\sum_{i=x,y}v\hbar\Pi_\mathrm{i}\tau_\mathrm{x}\sigma_\mathrm{i}+t_\mathrm{z}\sin(k_\mathrm{z}a)\tau_\mathrm{x}\sigma_\mathrm{z}\label{eq:Hamiltonian}
\end{equation}
with $v=\sqrt{v_\mathrm{x}v_\mathrm{y}}$ and $\Delta(k_\mathrm{{z}})=\Delta+2C(1-\cos k_\mathrm{{z}}a)$. $\boldsymbol{\Pi}=\mathbf{k}+e\mathbf{A}$
represents the kinematic momentum, $\mathbf{A}$ denotes the vector
potential components, and $a$ is the lattice constant along $z$-direction.
The Zeeman effect is encoded by $m=g_\mathrm{{z}}\mu_\mathrm{{B}}B/2$ with $g_\mathrm{{z}}=21.3$
the g-factor\textsuperscript{\citep{liu2016zeeman,jiang2017landau,sun2020large}} and
$\mu_\mathrm{{B}}=5.788\times10^{-2}\mathrm{meV}\cdot\mathrm{T}^{-1}$ the
Bohr magneton.The parameters in our model  (listed in the caption of Fig. \ref{fig:onlyZeeman}) are derived from first-principles calculations \textsuperscript{\citep{tang2019three}}, and are consistent with magneto-infrared spectroscopy measurements \textsuperscript{\citep{chen2015magnetoinfrared}}.The magnetic field has two primary effects: the
orbital effect ($\mathbf{A}$), which drives the formation of Landau
levels, and the Zeeman effect ($m$), which induces spin-dependent energy splitting.In the presence of a magnetic field,the eigenstates and eigenenergies can still be solved analytically,as demonstrated in Supplementary Note 1,where we also derive the corresponding Green's function.

For $m=0$ and $\mathbf{A}=0$, the Hamiltonian possesses time-reversal
symmetry, resulting in a vanishing Hall conductivity. For $m\ne0$
and $\mathbf{A}=0$, the energy spectrum takes the form: $\varepsilon_{\mathrm{s\zeta}}(\mathbf{k})=\zeta\sqrt{\mathcal{M}_\mathrm{{s}}^{2}+\hbar^{2}v^{2}\mathbf{k}_{\perp}^{2}}$
where $\mathbf{k}_{\perp}=(k_\mathrm{{x}},k_\mathrm{{y}})$, $\mathcal{M}_\mathrm{{s}}(k_\mathrm{z})=\Delta_{\parallel}(k_\mathrm{{z}})+sm$,
and $\Delta_{\parallel}(k_\mathrm{{z}})=\sqrt{\Delta^{2}(k_\mathrm{{z}})+t_\mathrm{{z}}^{2}\sin^{2}k_\mathrm{{z}}a}$.
Here, $s=\pm$ represents spin index, and $\zeta=\pm$ denotes the
conduction and valence bands, respectively. The energy spectra for
$|\mathbf{k}_{\perp}|=0$  are shown  in Fig. \ref{fig:onlyZeeman}. Zeeman splitting breaks the degeneracy
of the energy levels. The Hall conductivity can be understood in terms
of the nonzero Berry curvature of the occupied states, expressed as:
$\sigma_{\mathrm{xy}}^{\mathrm{S}}=\frac{e^{2}}{V\hbar}\sum_{\mathbf{k},\zeta,\mathrm{s}}f(\varepsilon_\mathrm{{s\zeta}}-\mu)\Omega_\mathrm{{z}}^\mathrm{{s\zeta}}$,
where the Berry curvature is given by $\Omega_\mathrm{{z}}^{\mathrm{s\zeta}}=-\frac{s\hbar^{2}v^{2}\mathcal{M}_\mathrm{{s}}}{2\varepsilon_\mathrm{{s\zeta}}^{3}}$ and $f$ is the Fermi-Dirac distribution.
The  momentum-space distribution of Berry curvature is visualized through color mapping of $\Omega_\mathrm{{z}}$  on the electronic band structure.
The superscript $\mathrm{S}$ indicates that this contribution arises
from quantum geometry effects due to spin (S)-splitting induced by
the magnetic field. When $|m|$ increases beyond the band gap $|\Delta|$,
a band crossing occurs, resulting in the creation of a pair of Weyl
nodes. Particularly when $\mu=0$, the Hall conductivity is given
by $\sigma_\mathrm{{xy}}^{\mathrm{S}}=\frac{e^{2}k_\mathrm{{c}}}{2\pi h}$, with $k_\mathrm{{c}}$
representing the distance between the two Weyl nodes\textsuperscript{\citep{burkov2014anomalous}}.
For higher Zeeman fields $|m|>|\Delta+4C|$, two Weyl points annihilate
at the Brillouin zone boundary, causing the system to become insulating
again. In this situation, the Hall conductivity becomes a constant
at $\sigma_\mathrm{{xy}}^{\mathrm{S}}=\frac{e^{2}}{ha}$ with $a$ as the lattice
constant in $z$-direction\textsuperscript{\citep{bernevig2007theory}}. The Hall conductivity
can also be calculated by using Kubo-streda formula \textsuperscript{\citep{streda1982theory,mahan2000many}},
which separates Hall conductivity into two distinct contributions,
$\sigma_\mathrm{{xy}}^{\mathrm{S}}=\sigma_\mathrm{{xy}}^{\mathrm{I}}+\sigma_{xy}^{\mathrm{II}}$
where $\sigma_\mathrm{{xy}}^{\mathrm{I}}$ describes the response at the Fermi
surface, and $\sigma_\mathrm{{xy}}^{\mathrm{II}}$ represents a nondissipative contribution from states below the Fermi
energy\textsuperscript{\citep{bastin1971quantum,stvreda1975galvanomagnetic,smrcka1977transport}}.
At zero temperature, these two components of the Hall conductivity
can be obtained as
\begin{align}
\sigma_\mathrm{{xy}}^{\mathrm{I}} & =\frac{e^{2}}{2h}\int_{-\pi/a}^{\pi/a}\frac{dk_\mathrm{{z}}}{2\pi}\sum_{s=\pm}s\Theta(\mu^{2}-\mathcal{M}_\mathrm{{s}}^{2})\frac{\mathcal{M}_\mathrm{{s}}}{|\mu|}\label{eq:sigmaxyI}
\end{align}
and
\begin{align}
\sigma_\mathrm{xy}^{\mathrm{II}}=\frac{e^{2}}{2h}\int_{-\pi/a}^{\pi/a}\frac{dk_{\mathrm{z}}}{2\pi}\sum_{\mathrm{s}=\pm}s\Theta(\mathcal{M}_\mathrm{s}^{2}-\mu^{2})\mathrm{sgn}(\mathcal{M}_\mathrm{{s}})\label{eq:sigmaxyII}
\end{align}
where $\mathrm{sgn}$ denotes the sign function, $\Theta$ is the
Heaviside step function, and $\mu$ is the chemical potential. The
Hall conductivity can be interpreted as a summation over each $k_{\mathrm{z}}$
slice of a two-dimensional system. We plot the chemical potential dependence of $\sigma_{\mathrm{xy}}^{\mathrm{I}}$, $\sigma_{\mathrm{xy}}^{\mathrm{II}}$, and $\sigma_{\mathrm{xy}}^{\mathrm{S}}$ in Fig. \ref{fig:onlyZeeman}.  Fig. \ref{fig:onlyZeeman}\textbf{a} demonstrates that before Weyl point formation, all three components vanish identically within the band gap. After the Weyl points emerge  (Fig. \ref{fig:onlyZeeman}\textbf{b}), $\sigma_\mathrm{{xy}}^{\mathrm{S}}$ generally receives contributions from both Fermi surface and Fermi sea terms. At charge neutrality, $\sigma_\mathrm{{xy}}^{\mathrm{S}}$ is solely determined by the Fermi sea contribution. For small magnetic field, both
$\sigma_{\mathrm{xy}}^{\mathrm{I}}$ and $\sigma_{\mathrm{xy}}^{\mathrm{II}}$ exhibit
a linear dependence on $B$ through the Zeeman term $m$. To first
order in the magnetic field expansion, the total Hall conductivity
can be obtained as $\sigma_{\mathrm{xy}}^{\mathrm{S}}=\frac{e^{2}}{2\pi h}\frac{k_{0}}{|\mu|}g_{\mathrm{z}}\mu_{B}B$,
where $k_{0}$ is the Fermi wavevector along the z-direction in the
absence of the Zeeman field, determined by $\Delta_{\parallel}(k_{0})=|\mu|$.

\begin{figure}
\includegraphics[width=12cm]{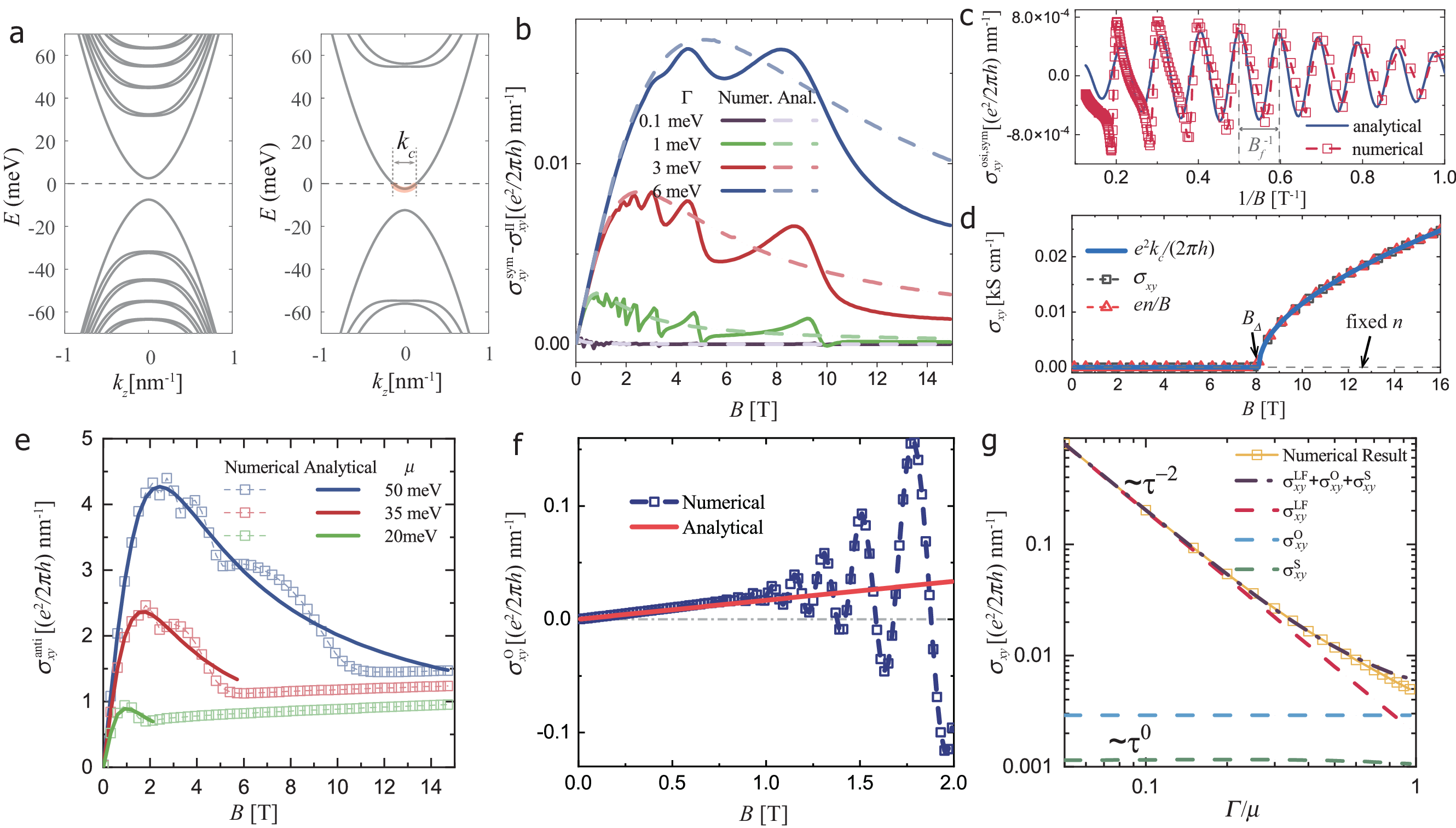}

\caption{\protect\label{fig:withOrbit}\textbf{Orbital Magnetic Effects on Hall Conductivity.}
\textbf{a} Landau level spectrum with orbital effects. The two lowest Landau levels $\varepsilon_{0\mathrm{s}}$ align precisely with the  spin-down states in the energy spectrum of  Fig. \ref{fig:onlyZeeman}. $k_\mathrm{c}$  denotes the Fermi wavevector of the lowest Landau level at charge neutrality $\mu=0$.
\textbf{b} Magnetic field dependence of the symmetric Hall conductivity $\sigma_\mathrm{xy}^{\mathrm{sym}}-\sigma_\mathrm{xy}^{\mathrm{II}}$  at fixed chemical potential $\mu=50$meV for different $\Gamma$. 
\textbf{c} Quantum oscillations in the symmetric Hall conductivity  $\sigma_\mathrm{xy}^\mathrm{osi,sym}$ : comparison between numerical calculations and analytical results for  $\mu=50$meV and $\Gamma=6$meV. The oscillation frequency field $B_\mathrm{f}\approx 10.74T$  corresponds to the extremal Fermi surface cross-section via the Onsager relation.
\textbf{d} The Hall conductivity $\sigma_\mathrm{{xy}}$ as a function of $B$ for $\mu=0$. The blue line marks the value $e^2k_\mathrm{c}/(2\pi h)$, corresponding to the separation of the Weyl nodes.  The black squares denote the numerical results for disorder broadening $\Gamma=0.1$ meV. The red triangles represent the result for $en/B$ with $n$ as the carrier density.
\textbf{e} Antisymmetric component of 
$\sigma_\mathrm{{xy}}$: numerical results compared with analytical expressions for different $\mu$ values with $\Gamma=3$meV.
\textbf{f} Comparison between numerical calculations and the analytical model for the orbital quantum correction $\sigma_\mathrm{{xy}}^{O}$ at $\mu=50$ meV and $\Gamma=3$meV.
\textbf{g} Analytical results for the Hall conductivity at low magnetic fields
($B=0.05$T): $\sigma_\mathrm{{xy}}^{\mathrm{LF}}$ (red dashed), $\sigma_\mathrm{{xy}}^{\mathrm{O}}$(blue
dashed), and $\sigma_\mathrm{xy}^{\mathrm{S}}$ (green dashed) as a function
of $\Gamma/\mu$. The purple line represents the summation of these
contributions. The yellow line with squares corresponds to the numerical
results. The chemical potential is $\mu=20$meV
in the simulations.}
\end{figure}

\paragraph*{Orbital Magnetic Effects Across Semiclassical and Quantum Oscillation Regimes}

We will now address the orbital effect ($m\ne0$ and $\mathbf{A}\ne0$).
The momentum perpendicular to the magnetic field is quantized as $\mathbf{k}_{\perp}^{2}\to2N/l_{B}^{2}$,
where $l_{B}=\sqrt{\hbar/eB}$ is the magnetic length, and $N=0,1,2,...$ labels the Landau levels. 
The eigenenergies become $\varepsilon_{0\mathrm{s}}=s\mathcal{M}_{-\mathrm{s}}$ for
the lowest Landau levels (LLLs) with $N=0$ 
and $\varepsilon_\mathrm{Ns\zeta}=\zeta\sqrt{\mathcal{M}_\mathrm{{s}}^{2}+N\eta^{2}}$
for higher Landau levels with $N\ge1$ (gray lines in Fig. \ref{fig:withOrbit}\textbf{a}), where
$\eta=\sqrt{2}v\hbar/l_\mathrm{B}$ represents the cyclotron energy. Due
to the presence of Zeeman field $m$, the LLLs are no longer symmetric
under electron-hole transformation, while the higher Landau levels
remain symmetric, although with broken degeneracy.  As demonstrated in the Supplementary Note 2, we rigorously derive the Hall magnetoconductivity ($\sigma_\mathrm{xy}$)
and transverse magnetoconductivity ($\sigma_\mathrm{xx}$) at arbitrary magnetic fields using the Kubo formula within the Landau level representation, neglecting vertex corrections. Disorder effects are incorporated by modeling
the Landau levels as Lorentzians with a constant broadening width
$\Gamma$, corresponding to a relaxation time $\tau=\hbar/(2\Gamma)$. These conductivities can be decomposed into antisymmetric ($\sigma_\mathrm{ab}^{\mathrm{anti}}$) and the symmetric ($\sigma_\mathrm{ab}^{\mathrm{sym}}$) terms: $\sigma_\mathrm{ab}=\sigma_\mathrm{ab}^{\mathrm{anti}}+\sigma_\mathrm{ab}^{\mathrm{sym}}$ based on symmetry considerations. The antisymmetric Hall conductivity changes sign under carrier type reversal ($\sigma_\mathrm{ab}^{\mathrm{anti}}(\mu)=-\sigma_\mathrm{ab}^{\mathrm{anti}}(-\mu)$), while the symmetric conductivity retains its sign ($\sigma_\mathrm{ab}^{\mathrm{sym}}(\mu)=\sigma_\mathrm{ab}^{\mathrm{sym}}(-\mu)$), where $a,b=x,y$.   Furthermore, the system exhibits three distinct regimes based on magnetic field strength as illustrated in Fig. \ref{fig:illustration_diagram}\textbf{c}: (i) Semiclassical regime
($B\le\chi^{-1}$ with $\chi=\frac{ev^{2}\tau}{\mu}$ represents
the mobility): The magnetic field is weak enough and disorder broadening smears the Landau levels. (ii) Quantum oscillations regimes: At higher magnetic fields, the system enters a regime characterized by quantum oscillations. (iii)  Quantum limit ($B\ge\frac{\mu^{2}-\Delta^{2}}{2\hbar v^{2}e}$): In this regime, the magnetic field is strong enough that only the lowest Landau level (LLL) is partially filled.

Using a small magnetic field expansion of the orbital magnetic effects and fully incorporating the Zeeman-induced band splitting, we derive the magnetoconductivities in the semiclassical regime while simultaneously capturing the background contributions that persist into the quantum oscillation regime at higher fields. The symmetric  Hall conductivity $\sigma_\mathrm{xy}^{\mathrm{sym}}$ is given
by:
\begin{align}
\sigma_{\mathrm{xy}}^{\mathrm{sym}}=	\frac{\sigma_\mathrm{xy}^{\mathrm{I}}}{1+\chi^{2}B^{2}}+\sigma_\mathrm{xy}^\mathrm{II}\label{eq:anomalous_hall_A_sym}.
\end{align}
The symmetric Hall conductivity arises from the Zeeman splitting-induced Berry curvature effect, though it is modified by orbital effects. This analytic expression is numerically validated in Supplementary Note 3. For sufficient weak fields ($\chi B\ll 1$), where orbital effects become negligible,  $\sigma_\mathrm{xy}^{\mathrm{sym}}\simeq\sigma_\mathrm{xy}^{\mathrm{S}}$ exhibits linear field dependence. At stronger fields, orbital effects suppress Fermi surface contribution $\sigma_\mathrm{xy}^\mathrm{I}$, while preserving Fermi sea contribution $\sigma_\mathrm{xy}^\mathrm{II}$ - the topological component representing nondissipative contributions from states below the Fermi energy. Figure \ref{fig:withOrbit}\textbf{b}  shows the difference between the symmetric Hall conductivity and the Fermi sea contribution, $\sigma_\mathrm{xy}^{\mathrm{sym}}-\sigma_\mathrm{xy}^\mathrm{II}$, plotted as a function of $B$ for various scattering rates $\Gamma$. The difference initially increases linearly before decreasing with magnetic field, in excellent agreement with Eq. \ref{eq:anomalous_hall_A_sym}. At small $\Gamma$  (purple line), orbital effects completely suppress $\sigma_\mathrm{xy}^\mathrm{I}$, causing $\sigma_\mathrm{xy}^{\mathrm{sym}}$ to converge with $\sigma_\mathrm{xy}^\mathrm{II}$.  By subtracting the background contribution, we isolate the quantum oscillations in the symmetric Hall conductivity:

\begin{align}
\sigma_\mathrm{xy}^{\mathrm{sym},\mathrm{osc}}	\approx\frac{e^{2}}{\sqrt{2\pi} h}\sum_\mathrm{s}\frac{s\Theta[\mu^{2}-\mathcal{M}_\mathrm{s}^2(0)]\mathcal{M}_\mathrm{s}(0)}{|\mu|(1+\chi^2 B^2)}\frac{\cos(S_\mathrm{s}^{(0)}l_\mathrm{B}^{2}-s\frac{\pi}{4})}{|S_\mathrm{s}^{(2)}l_\mathrm{B}^{2}|^{1/2}}\exp(-\frac{\pi}{|\chi B|})\label{eq:anomalous_hall_osi}
\end{align}
where $S_\mathrm{s}^{(0)}$ and $S_\mathrm{s}^{(2)}$ are the zeroth and second order coefficients in the $k_\mathrm{z}$ expansion of $S_\mathrm{s}=\pi(\mu^2-\mathcal{M}_\mathrm{s}^2)/(\hbar^2v^2)\approx S_\mathrm{s}^{(0)}+\frac{1}{2}S_\mathrm{s}^{(2)}k_\mathrm{z}^2$. The quantum oscillation fully comes from the Fermi surface contribution. The two bands $s=\pm$ produce two distinct Fermi surfaces as a result of Zeeman splitting, giving rise to a beating pattern in quantum oscillations due to their slightly different frequencies. At small magnetic fields, the splitting is minimal, and the symmetric component of the oscillatory Hall conductivity follows the form $\sigma_\mathrm{xy}^{\mathrm{sym},\mathrm{osc}}\sim \cos(2\pi B_\mathrm{f}/B)$ with $B_\mathrm{f}\approx(\mu^2-\Delta^2)/(2e\hbar v^2)\approx 10.47T$.   As shown in \ref{fig:withOrbit}\textbf{c}, our numerical results show good agreement with the analytical expression in Eq. \ref{eq:anomalous_hall_osi}.  
The relation $\sigma_\mathrm{xy}^{\mathrm{sym}}\simeq\sigma_\mathrm{xy}^{\mathrm{II}}$ remains valid even in the quantum limit. This is demonstrated in Fig.  \ref{fig:withOrbit}\textbf{d} for the case $\mu=0$, where an infinitesimal magnetic field drives the system into the quantum limit. Here, the Hall conductivity exhibits perfect particle-hole symmetry ($\sigma_\mathrm{xy}=\sigma_\mathrm{xy}^{\mathrm{sym}}$) and becomes nonzero for $B>B_{\Delta}$, coinciding with the crossing of the zeroth Landau level through $\mu=0$ (Fig. \ref{fig:withOrbit}\textbf{a}). In this regime, $\sigma_\mathrm{xy}=\sigma_\mathrm{xy}^{\mathrm{II}}=\frac{e^{2}k_\mathrm{c}}{2\pi h}$, as shown by the blue line.

The antisymmetric Hall conductivity can be expressed as the sum of two main contributions:
\begin{align}
\sigma_\mathrm{xy}^{\mathrm{anti}}= & \frac{\sigma_\mathrm{xy}^{\mathrm{LF}}+\sigma_\mathrm{xy}^{\mathrm{O}}}{1+\chi^{2}B^{2}}\label{eq:anomalous_hall_A_anti}
\end{align}
 where $\sigma_\mathrm{xy}^{\mathrm{LF}}$ corresponds to the classical Lorentz
force response, and $\sigma_\mathrm{xy}^{\mathrm{O}}$ incorporates the correction
from the quantum metric, orbital magnetization, and other orbital
(O) field-induced effects.  Figure \ref{fig:withOrbit}\textbf{e} compares numerical calculations with analytical results for the antisymmetric Hall conductivity $\sigma_\mathrm{xy}^{\mathrm{anti}}$ as a function of magnetic field $B$ at  different chemical potential $\mu$.
The analytical expressions not only match the numerical results in the classical regime but also correctly reproduce the background behavior that persists into the quantum oscillation regime at higher fields. As shown in Fig. \ref{fig:withOrbit}\textbf{f}, after subtracting the classical Lorentz force contribution,  $\sigma_\mathrm{xy}^{\mathrm{O}}$ exhibits a linear field dependence, clearly revealing the quantum corrections arising from orbital magnetic effects. By retaining terms
to linear order in the magnetic field, the explicit contributions
of different mechanisms to $\sigma_\mathrm{xy}$ are listed in Table I. We emphasize that our analytical expressions are derived in the weak scattering limit $\Gamma\to 0$.
When $\Gamma$ is not small, this decomposition breaks down, but the full Kubo-Streda formula - evaluated using the complete Green's function without expansion - remains rigorously valid. Fig.
\ref{fig:withOrbit}\textbf{g} shows the contributions to the Hall
conductivity as a function of $\Gamma/\mu$ for a small magnetic field.
$\sigma_\mathrm{xy}^{\mathrm{S}}$ (green dashed line) and $\sigma_\mathrm{xy}^{\mathrm{O}}$
(blue dashed line) from quantum geometric effects are independent
of $\tau$ while the classical contribution $\sigma_\mathrm{xy}^{\mathrm{LF}}$
(red dashed line) varies as $\tau^{-2}$. As $\Gamma/|\mu|$ approaches
$1$, the quantum geometric correction grows comparable to the classical Lorentz force contribution, driving a crossover in the total response from $\tau^{-2}$ to $\tau^{0}$ scaling. The analytical results (purple line) accurately describe
this crossover behavior, as confirmed by precise numerical results
(yellow line with squares).

To analyze the magnetoresistance, we need to obtain the transverse
magnetoconductivity $\sigma_\mathrm{xx}$. The transverse magnetoconductivity
is given by: $\sigma_\mathrm{xx}=\frac{\sigma_\mathrm{xx}^{0}}{1+\chi^{2}B^{2}},$
where $\sigma_\mathrm{xx}^{0}$ includes both electron-hole incoherent ($\sigma_{\mathrm{xx},\mathrm{in}}^{0}$)
and coherent ($\sigma_\mathrm{xx,\mathrm{co}}^{0}$) contributions (See Supplementary Note 4 for the explicit forms of these terms).
The incoherent contribution arises from the retarded-advanced channel
in Kubo formula, while the coherent contribution originates from the
retarded-retarded channel. As $\sigma_{\mathrm{xx},\mathrm{co}}^{0}/\sigma_{\mathrm{xx},\mathrm{in}}^{0}\sim\Gamma/\mu$,
electron-hole coherence plays an important role near the band bottom
($\Gamma\sim|\mu|$).

\begin{figure}
\includegraphics[width=12cm]{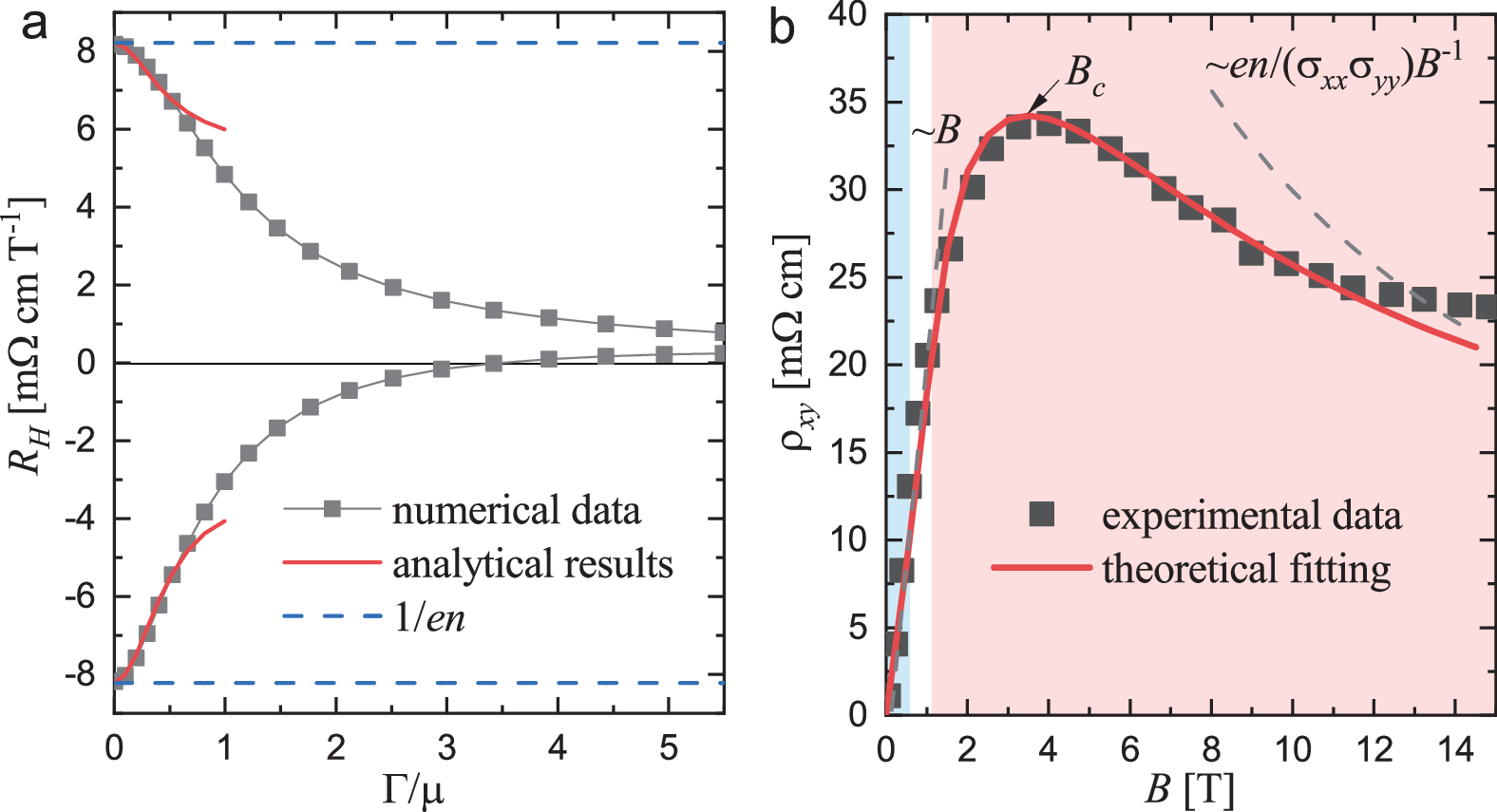}\caption{\protect\label{fig:(a)-Hall-coefficient}\textbf{Quantum Geometric Renormalization of Hall coefficient and Nonlinear Hall Resistivity.} \textbf{a} Hall coefficient $R_{H}$
versus $\Gamma$ for a fixed carrier density $n=\pm8\times10^{16}\,\mathrm{cm^{-3}}$
in  the semiclassical regime, for both electron- and hole-type carriers.
The blue dashed lines represent the classical result $1/en$. The
red lines show the analytical results derived from Eq. (\ref{eq:R_H}), while the gray squares correspond to numerical solutions obtained
from the full expression. \textbf{b} Comparison between experimental data
(black dashed line with squares) from Ref.[38] 
and theoretical simulations (red line) based on our model. The simulations
use fitting parameters $n=3\times10^{16}\mathrm{cm^{-3}}$, $\Gamma=6.5\,\mathrm{meV}$,
$\Delta=5\,\mathrm{meV}$, with other model parameters consistent
with previous simulations. The blue and red shaded regions indicate
the semiclassical regime and quantum limit, respectively. Dashed lines
serve as eye guides for $\sim B$ and $en/(\sigma_\mathrm{xx}\sigma_\mathrm{yy})B^{-1}$.
$B_\mathrm{c}\simeq3.8\,\mathrm{T}$ indicates the critical field.}
\end{figure}

In the semiclassical regime, the Hall resistivity $\rho_\mathrm{xy}\simeq\sigma_\mathrm{xy}/(\sigma_\mathrm{xx}\sigma_\mathrm{yy})=R_\mathrm{H}B$
remains linear in $B$, with the Hall coefficient $R_\mathrm{H}=\partial\rho_\mathrm{xy}/\partial B$
given by:
\begin{equation}
R_\mathrm{H}=\frac{1}{B}\frac{\sigma_{\mathrm{xy}}^{\mathrm{LF}}+\sigma_\mathrm{xy}^{\mathrm{O}}+\sigma_\mathrm{xy}^{\mathrm{S}}}{(\sigma_{\mathrm{xx},\mathrm{in}}^{0}+\sigma_{\mathrm{xx},\mathrm{co}}^{0})^{2}}.\label{eq:R_H}
\end{equation}
When $\Gamma/|\mu|\ll1$, $\sigma_{\mathrm{xx},\mathrm{in}}^{0}\simeq\sigma_{0}\gg\sigma_{\mathrm{xx},\mathrm{co}}^{0}$
and Lorentz force contribution $\sigma_\mathrm{xy}^{\mathrm{LF}}\simeq\chi B\sigma_{0}$
dominates $\sigma_\mathrm{xy}$, then $R_\mathrm{H}\simeq1/en$ reduces to the classical
result. When $\Gamma\sim|\mu|$, the contribution from quantum geometric
effects and electron-hole coherence can no longer be neglected. Consequently,
the Hall coefficient must be modified after accounting for these effects.
As shown in Fig. \ref{fig:(a)-Hall-coefficient}\textbf{a}, we plot $R_\mathrm{H}$
as a function of $\Gamma/\mu$ for a fixed carrier density $n$, considering
both electron- and hole-type carriers. As $\Gamma$ increases, $R_\mathrm{H}$
gradually deviates from the classical result $1/en$. The analytical
solution (red lines) from Table 1, derived through a $\Gamma$ expansion,
agrees well with numerical calculations for  $\Gamma/\mu<1$. However,
in the $\Gamma/\mu>1$ regime, higher-order quantum geometric corrections
($\propto\Gamma^{2}...$) become significant and must be included
to accurate description. The conductivity components exhibit distinct
symmetry properties: $\sigma_\mathrm{xy}^{\mathrm{LF}}$ and $\sigma_\mathrm{xy}^{\mathrm{O}}$
are antisymmetric under carrier-type reversal ($\mu\to-\mu$), while
$\sigma_\mathrm{xy}^{\mathrm{S}}$ is symmetric. This leads to markedly different
behavior for two carrier types-with increasing $\Gamma$, the hole-type
$R_\mathrm{H}$ can vanish or even change sign when $\sigma_\mathrm{xy}^{\mathrm{S}}$
compensates or dominates the $\sigma_\mathrm{xy}^{\mathrm{LF}}$ and $\sigma_\mathrm{xy}^{\mathrm{O}}$,
whereas the electron-type $R_\mathrm{H}$ maintains its original sign throughout.
By analyzing the frequency of the Shubnikov--de Haas (SdH) oscillations,
which is directly proportional to the cross-sectional area of the
Fermi surface, the carrier density of the system can be determined
\textsuperscript{\citep{pippard1989magnetoresistance}}. By comparing the Hall coefficients
obtained from Hall resistivity measurements with the carrier density
extracted from SdH oscillations, one can identify the field-induced
unconventional Hall effect in experiments. This approach provides
a robust method to distinguish between conventional Hall effects and
anomalous contributions arising from quantum geometry in the out-of-plane
magnetic field configuration.

\begin{table}
\begin{tabular}{cccc}
\toprule 
 & \multicolumn{2}{c}{semiclassical regime} & quantum limit\tabularnewline
\midrule
\midrule 
\multirow{3}{*}{$\sigma_\mathrm{xy}$} & Lorentz force ($\sigma_\mathrm{xy}^{\mathrm{LF}}$) & $\frac{e^{3}v^{2}\langle v_\mathrm{x}^{2}\rangle\tau^{2}\nu_\mathrm{\mu}}{\mu}B$ & \multirow{3}{*}{$\frac{en}{B}$}\tabularnewline
\cmidrule{2-3}
 & orbital effect ($\sigma_\mathrm{xy}^{\mathrm{O}}$) & $\frac{e^{3}v^{4}\hbar^{2}\nu_\mathrm{\mu}}{8\mu^{3}}B$ & \tabularnewline
\cmidrule{2-3}
 & spin effect ($\sigma_\mathrm{xy}^{\mathrm{S}}$) & $\frac{e^{2}}{2\pi h}\frac{k_{0}}{|\mu|}g_\mathrm{z}\mu_\mathrm{B}B$ & \tabularnewline
\midrule 
\multirow{2}{*}{$\sigma_\mathrm{xx}$} & incoherent ($\sigma_{\mathrm{xx},\mathrm{in}}^{0}$) & $e^{2}\langle v_\mathrm{x}^{2}\rangle\tau\nu_\mathrm{\mu}$ & \multirow{2}{*}{$\frac{e^{2}}{2\pi ha}\frac{\Gamma}{t_\mathrm{z}}\mathcal{F}$}\tabularnewline
\cmidrule{2-3}
 &  coherent ($\sigma_{\mathrm{xx},\mathrm{co}}^{0}$) & $\frac{e^{2}}{\pi ha}$ & \tabularnewline
\bottomrule
\end{tabular}

\caption{\textbf{The contributions to the Hall and transverse conductivity in
the semiclassical regime and quantum limit.} $\sigma_{\mathrm{xy}}^{\mathrm{LF}}$ is Lorentz force contribution. $\sigma_{\mathrm{xy}}^{\mathrm{O}}$ is the orbital field-induced conductivity. $\sigma_{\mathrm{xy}}^{\mathrm{S}}$ is the spin-splitting-induced conductivity. $\sigma_{\mathrm{xx,in}}^{0}$ represents the electron-hole incoherent contribution to the transverse magnetoconductivity. $\sigma_{\mathrm{xx,co}}^{0}$ is the electron-hole coherent contribution to the transverse magnetoconductivity. $\nu_{\mu}=\frac{|\mu|}{2\pi\hbar^{2}v^{2}}\protect\int\frac{dk_\mathrm{z}}{2\pi}\sum_{\mathrm{s}=\pm}\Theta(\mu^{2}-\mathcal{M}_\mathrm{s}^{2})$
is the density of states at the Fermi energy. $\langle v_{\mathrm{x}}^{2}\rangle=\frac{1}{\nu_{\mu}}\protect\int[d\mathbf{k}]\sum_\mathrm{s,\zeta}(v_\mathrm{x,s\zeta})^{2}\delta(\mu-\varepsilon_\mathrm{s\zeta})$
represents the average square of the Fermi velocity over the Fermi
surface, where $v_{\mathrm{x,s\zeta}}=\frac{1}{\hbar}\partial_{k_\mathrm{x}}\varepsilon_\mathrm{s\zeta}$
is the Fermi velocity along $x$-direction. The Lorentz force term
can be rewritten in a more familiar form $\sigma_{\mathrm{xy}}^{\mathrm{LF}}=\chi B\sigma_{0}$,
where $\sigma_{0}=e^{2}D\nu_{\mu}\langle3v_\mathrm{x}^{2}/v^{2}\rangle$
with $D=\tau v^{2}/3$ as the diffusion constant.}
\end{table}

From a semiclassical perspective, the magnetic field modifies the conductivity through corrections to both the Berry curvature and band
energy of electronic states\textsuperscript{\citep{gao2014field,gao2019semiclassical,xiao2021thermoelectric,wang2024orbital}}. Using a weak magnetic field expansion of the Green's function (See the Methods section and Supplementary Note 6 for details),
we derive $\tau$-independent Hall conductivity $\sigma_\mathrm{xy}^{(0)}=\sigma_\mathrm{xy}^{O}+\sigma_\mathrm{xy}^{S}$ in terms of quantum geometric quantities. We find $\sigma_\mathrm{ab}^{(0)}=\sigma_\mathrm{ab}^{(0),\mathrm{wp}}+\sigma_\mathrm{ab}^{(0)\prime}$, where $\sigma_\mathrm{ab}^{(0),\mathrm{wp}}$ represents the previously known contribution derived from semiclassical wavepacket theory\textsuperscript{\citep{gao2014field,gao2019semiclassical,xiao2021thermoelectric,wang2024orbital}}, given by $\sigma_\mathrm{ab}^{(0),\mathrm{wp}}=\frac{e^{2}}{\hbar}\int[d\mathbf{p}][(-\frac{\partial f}{\partial\varepsilon})(B_\mathrm{c}F_\mathrm{cb}V_\mathrm{a}+\frac{1}{2}\epsilon_\mathrm{abc}\Omega_{c}(\mathbf{m}\cdot\mathbf{B})-(a\longleftrightarrow b)]$.
Here, $a,b,c$ denote Cartesian components (with Einstein summation convention), $\epsilon_\mathrm{abc}$ is
the Levi-Civita symbol,  $F_\mathrm{cb}$ is the anomalous orbital polarizability
(AOP), $\Omega_\mathrm{c}$  the Berry curvature, and $\mathbf{m}$  the intraband
orbital magnetic moment. The additional term $\sigma_\mathrm{ab}^{(0)\prime}$ is a new correction obtained through Green's function techniques,  which contains higher-order energy derivatives of the Fermi-Dirac distribution:
\begin{align}
\sigma_\mathrm{ab}^{(0)\prime}= & \frac{e^{2}}{\hbar}\int[d\mathbf{p}]\{\frac{\partial f}{\partial\epsilon}\frac{1}{2}g_\mathrm{ac}(\mathbf{B}\times\nabla_{\mathbf{p}}V_\mathrm{b})_\mathrm{c}-\frac{\partial^{3}f}{\partial\epsilon^{3}}\frac{1}{8}n_\mathrm{ac}V_\mathrm{b}(\mathbf{B}\times\mathbf{V})_{c}\nonumber \\
 & -\frac{\partial^{2}f}{\partial\epsilon^{2}}\frac{1}{8\mu}\mathrm{Im}\langle[\hat{V}_\mathrm{a},\hat{V}_\mathrm{b}]\rangle(\mathbf{m}\cdot\mathbf{B})]-(a\longleftrightarrow b)\},\label{eq:classical_correction}
\end{align}
where $g_\mathrm{ac}$ is the quantum metric tensor,  and $n_\mathrm{ab}=\mathrm{Re}\langle\partial_{p_\mathrm{a}}u_{\alpha\mathbf{p}}|(\varepsilon_{\alpha}-\hat{H}_{0})|\partial_{p_\mathrm{b}}u_{\alpha\mathbf{p}}\rangle$
with $\epsilon_{\alpha}$ and $|u_{\alpha\mathbf{p}}\rangle$ as the
eigenenergies and eigenstates for the band intersecting the Fermi
surface. These terms are non-vanishing and
essential for recovering the complete expression for  $\sigma_\mathrm{xy}^{\mathrm{O}}$.

\paragraph*{Nonlinear Hall resistivity in quantum limit}

In the quantum limit, disorder has minimal impact on $\sigma_\mathrm{xy}$,
allowing us to neglect impurity effects when evaluating $\sigma_\mathrm{xy}$.
As shown in Supplementary Note 5, the total Hall conductivity can be rigorously shown to satisfy $\sigma_\mathrm{xy}=\frac{en}{B}$,
where $n$ is the carrier density, with the charge neutrality point
defined as the midpoint between the two LLLs(see Supplementary Note 7). As demonstrated in Fig. \ref{fig:withOrbit}\textbf{d}, we validate this relation by examining the case where  $\mu=0$. Here, an infinitesimal magnetic field is sufficient to drive the system into the quantum limit. The Hall conductivity exhibits perfect particle-hole symmetry and develops a nonzero value for $B>B_{\Delta}$. Our numerical results for the Hall conductivity (black squares) show excellent agreement with the analytical expression (red triangles), even when Weyl nodes form without orbital effects. This agreement persists because the carrier density can be expressed as $n=eBk_\mathrm{c}/(2\pi h)$. If the carrier density is fixed at the charge neutrality point (n=0), the Hall conductivity vanishes identically (black dashed line).  As illustrated by Fig. \ref{fig:illustration_diagram}\textbf{c}, the leading-order conductivity $\sigma_\mathrm{xx}$
arises from the inter-band velocity and the scatterings between the
0th bands with the bands of 1, which are higher-order perturbation
effects\textsuperscript{\citep{zhang2016linear}}. After calculation, we find: $\sigma_\mathrm{xx}=\frac{e^{2}}{2\pi ha}\frac{\Gamma}{t_\mathrm{z}}\mathcal{F}$
where $\mathcal{F}$ is a dimensionless integral of order unity, weakly
dependent on $m/\Gamma$ and $\mu/\Gamma$. For estimating $\rho_\mathrm{xy}$,
$\mathcal{F}$ can be approximated as a constant.  $\sigma_\mathrm{xx}$
is proportional to $\Gamma$, while $\sigma_\mathrm{xy}$ is inversely proportional
to $B$. This leads to a field-driven competition between these two
conductivity components. At small field, where $\sigma_\mathrm{xy}^{2}\gg\sigma_\mathrm{xx}\sigma_\mathrm{yy}$,
$\rho_\mathrm{xy}$ exhibits a classical linear dependence: $\rho_\mathrm{xy}\simeq B/en$.
Conversely, at larger fields, where $\sigma_\mathrm{xy}^{2}\ll\sigma_\mathrm{xx}\sigma_\mathrm{yy}$,
$\rho_\mathrm{xy}$ becomes inversely proportional to $B$: $\rho_\mathrm{xy}\simeq\frac{en}{\sigma_\mathrm{xx}\sigma_\mathrm{yy}}B^{-1}$.
The transition between these two regimes occurs at a critical magnetic
field:
\begin{equation}
B_\mathrm{c}=\frac{t_\mathrm{z}}{\Gamma}\frac{2\pi ha}{e\mathcal{F}}n.\label{eq:B_c}
\end{equation}
For larger values of $\Gamma$ and smaller carrier density $n$, $B_\mathrm{c}$
decreases, making the transition easier to observe. We apply our theory
to explain the experimental data from Ref. [38].
As shown in Fig. \ref{fig:(a)-Hall-coefficient}\textbf{b}, for a low carrier
density $n=3\times10^{16}\mathrm{cm^{-3}}$, the system enters the
quantum limit at around $B_\mathrm{QL}\approx1\,\mathrm{T}$, indicated by
the red shaded region. Quantum oscillations emerge in the regime where $\chi^{-1}<B<B_\mathrm{QL}=(\mu^2-\Delta^2)/(2\hbar v^2e)$, as illustrated in Fig.\ref{fig:illustration_diagram}. However, when the mobility  is relatively low and the carrier density is small, the quantum oscillations in the intermediate field range become difficult to observe.Using $\mathcal{F}=0.6$ and the band gap and
broadening parameters provided in Ref. [38],
the critical magnetic field is estimated to be $B_\mathrm{c}\approx3.8\,\mathrm{T}$.
Our theory accurately explains the experimental results. Additionally, it self-consistently reproduces the longitudinal conductivity, as demonstrated in Supplementary Note 8.

\section{Conclusion}\label{sec12}

In conclusion, we propose a quantum theory for the unconventional
Hall effect in paramagnetic Dirac materials, combining intrinsic band
topology with field-induced quantum effects of LLs. By systematically
separating the total Hall conductivity contributions based on symmetry
and physical origin, we clarify how each component evolves under an
applied magnetic field. In the semiclassical regime, quantum geometry
effects modify the Hall coefficient, while in the quantum limit, a
competition between transverse and Hall conductivities leads to the
nonmonotonic field dependence of the Hall effect. This work provides
a comprehensive framework for understanding the intricate interplay
between quantum geometry, magnetic field, and disorder in Dirac materials,
and also offers valuable insights into the origin of anomalous transport
phenomena observed in other nonmagnetic Dirac materials, such as $\mathrm{Cd}_{3}\mathrm{As}_{2}$\textsuperscript{\citep{liang2017anomalous,nishihaya2025anomalous}}.

At last, we emphasize the fundamental distinction between our proposed mechanism and conventional multiband models. The origin of the unconventional Hall effect in our framework stems from quantum geometric effects that persist regardless of Fermi surface topology or temperature, whereas multiband models require specific band structure conditions and thermally activated carriers. If angle-resolved photoemission spectroscopy (ARPES) reveals a single Fermi surface (exclusively electron- or hole-like) without additional bands crossing the Fermi level, this would rule out multiband contributions to the unconventional Hall effect. In such a scenario, the unconventional Hall response observed at low temperatures ($k_\mathrm{B}T \ll \Delta$) must arise from intrinsic quantum geometric effects, as thermally excited minority carriers are exponentially suppressed. This distinction is particularly crucial in systems like $\mathrm{ZrTe}_5$, where the bandgap and Fermi level positioning are sensitive to sample stoichiometry and external perturbations. Transport measurements under controlled doping or strain could thus serve as additional tests to isolate the geometric contribution.

\section{Methods}\label{sec11}

\paragraph*{Eigensolutions and Green's function for the Dirac Hamiltonian in
a finite magnetic field}

We perform a unitary transformation to Hamiltonian  [Eq. (\ref{eq:Hamiltonian})], 
$U=e^{i\tau_\mathrm{y}\sigma_{z}\frac{\theta}{2}}$ with $\theta=\arctan\frac{t_\mathrm{z}\sin k_\mathrm{z}}{\Delta(k_\mathrm{z})}$,
leading to
\begin{align*}
H^{\prime} & =UHU^{-1}\\
 & =\left[\begin{array}{cccc}
\Delta_{\parallel}+m & 0 & 0 & \eta a\\
0 & \Delta_{\parallel}-m & \eta a^{\dagger} & 0\\
0 & \eta a & -\Delta_{\parallel}+m & 0\\
\eta a^{\dagger} & 0 & 0 & -\Delta_{\parallel}-m
\end{array}\right]
\end{align*}
where the ladder operators are defined as $a=\frac{v}{\eta}\Pi_\mathrm{x}-i\frac{v}{\eta}\Pi_\mathrm{y}$
and $a^{\dagger}=\frac{v}{\eta}\Pi_\mathrm{x}+i\frac{v}{\eta}\Pi_\mathrm{y}$ ,
and $\Delta_{\parallel}(k_\mathrm{z})=\sqrt{\Delta(k_{z})^{2}+(t_\mathrm{z}\sin k_\mathrm{z})^{2}}$.
The Hamiltonian $H^{\prime}$ separates into two subblocks:
$H^{\prime}=H_{+}\oplus H_{-}$
 with 
$
H_{+}=\mathcal{M}_{+}\sigma_\mathrm{z}+\eta a\sigma_{+}+\eta a^{\dagger}\sigma_{-},H_{-}=\mathcal{M}_{-}\sigma_\mathrm{z}+\eta a^{\dagger}\sigma_{+}+\eta a\sigma_{-}
$
and $\mathcal{M}_{\pm}(k_\mathrm{z})=\Delta_{\parallel}\pm m$. The current
operator is obtained from $j_\mathrm{a}=ei\hbar^{-1}[H,r_\mathrm{a}]$ with $a=x,y$
denoting the coordinates in the plane perpendicular to the magnetic
field. In the subspace $s$ with $s=\pm$, the current operators are
$j_\mathrm{x}^\mathrm{s}=ev_\mathrm{x}\sigma_\mathrm{x}$ and $j_\mathrm{y}^\mathrm{s}=sev_\mathrm{y}\sigma_\mathrm{y}$.
The eigenfunctions of the $2\times2$ Hamiltonian are given by
\begin{align}
\psi_\mathrm{ns\zeta}(\mathbf{r})=\langle\mathbf{r}|\psi_\mathrm{ns\zeta k_\mathrm{x}k_\mathrm{z}}\rangle & =\frac{e^{ik_\mathrm{x}x+ik_\mathrm{z}z}}{\sqrt{L_\mathrm{x}L_\mathrm{z}}}\left[\begin{array}{c}
\zeta\cos\frac{\varphi_\mathrm{s\zeta n}}{2}\phi_{n-\frac{1+\mathrm{s}}{2}}(\xi)\\
\sin\frac{\varphi_\mathrm{s\zeta n}}{2}\phi_{n-\frac{1-\mathrm{s}}{2}}(\xi)
\end{array}\right]\label{eq:wave_function_B}
\end{align}
with $\xi=y/l+k_\mathrm{x}l$ , $\cos\varphi_\mathrm{ns\zeta}=\zeta\mathcal{M}_\mathrm{s}/\varepsilon_\mathrm{ns}$,
and $\sin\varphi_\mathrm{s\zeta n}=\sqrt{n}\eta/\varepsilon_\mathrm{ns}$. $\phi_\mathrm{n}(\xi)=\frac{e^{-\xi^{2}/2}}{2^{n/2}\pi^{1/4}\sqrt{n!l_\mathrm{B}}}H_\mathrm{n}(\xi)$ where $H_\mathrm{n}$ denotes the nth Hermite polynomial. The corresponding
eigenenergies are $\zeta\varepsilon_\mathrm{ns}=\zeta\sqrt{\mathcal{M}_\mathrm{s}^{2}+n\eta^{2}}$.
The wavefunctions for the lowest Landau levels are
\begin{align}
\psi_{0+}(\mathbf{r})=\langle\mathbf{r}|\psi_{0+k_\mathrm{x}k_\mathrm{z}}\rangle & =\frac{e^{ik_\mathrm{x}x+ik_\mathrm{z}z}}{\sqrt{L_\mathrm{x}L_\mathrm{z}}}\left[\begin{array}{c}
0\\
\phi_{0}(\xi)
\end{array}\right],\nonumber \\
\psi_{0-}(\mathbf{r})=\langle\mathbf{r}|\psi_{0-k_\mathrm{x}k_\mathrm{z}}\rangle & =\frac{e^{ik_\mathrm{x}x+ik_\mathrm{z}z}}{\sqrt{L_\mathrm{x}L_\mathrm{z}}}\left[\begin{array}{c}
\phi_{0}(\xi)\\
0
\end{array}\right],\label{eq:wave_function_n=00003D0}
\end{align}
with the corresponding eigenenergies given by $\varepsilon_{0s}=-s\mathcal{M}_{s}$.Then,
we can derive the retarded ($R$) and advanced $(A)$ Green's functions
\begin{align*}
G^\mathrm{R,A}(\omega;\mathbf{r},\mathbf{r}^{\prime})= & \sum_\mathrm{s}\frac{\langle\mathbf{r}|\psi_{0\mathrm{s}k_\mathrm{x}k_\mathrm{z}}\rangle\langle\psi_{0\mathrm{s}k_\mathrm{x}k_\mathrm{z}}|\mathbf{r}^{\prime}\rangle}{\omega^\mathrm{R,A}-\varepsilon_{0\mathrm{s}}}+\sum_{\mathrm{s},\zeta,\mathrm{n}\ge1}\frac{\langle\mathbf{r}|\psi_{\mathrm{ns}\zeta k_\mathrm{x}k_\mathrm{z}}\rangle\langle\psi_{\mathrm{ns}\zeta k_\mathrm{x}k_\mathrm{z}}|\mathbf{r}^{\prime}\rangle}{\omega^\mathrm{R,A}-\varepsilon_{\mathrm{ns}\zeta}}
\end{align*}
with $\omega^\mathrm{R,A}=\omega\pm i\Gamma$. By substituting Eqs. (\ref{eq:wave_function_B}) and (\ref{eq:wave_function_n=00003D0}),
we can obtain
\[
G(\omega;\mathbf{r},\mathbf{r}^{\prime})=e^{i\Phi(\mathbf{r}_{\perp},\mathbf{r}_{\perp}^{\prime})}\widetilde{G}(\omega;\mathbf{r}-\mathbf{r}^{\prime})
\]
where the Schwinger phase as $\Phi(\mathbf{r}_{\perp},\mathbf{r}_{\perp}^{\prime})=\int_{\mathbf{r}_{\perp}}^{\mathbf{r}_{\perp}^{\prime}}d\mathbf{r}_{\perp}\cdot\mathbf{A}(\mathbf{r}_{\perp})=-\frac{(x-x^{\prime})(y+y^{\prime})}{2l_\mathrm{B}^{2}}$
with $\mathbf{r}_{\perp}=(x,y)$ which breaks the translational invariance
explicitly and the translational invariant part is\textsuperscript{\citep{gusynin2006transport,tsaran2016magnetic}}
\begin{align*}
\widetilde{G}^\mathrm{R,A}(\omega;\mathbf{r}_{\perp}-\mathbf{r}_{\perp}^{\prime})
= & \frac{e^{ik_\mathrm{z}(z-z^{\prime})}}{2\pi l_\mathrm{B}^{2}}e^{-\frac{|\mathbf{r}_{\perp}-\mathbf{r}_{\perp}^{\prime}|^{2}}{4l_\mathrm{B}^{2}}}\Big\{\frac{\omega^\mathrm{R,A}-s\mathcal{M}_\mathrm{s}}{(\omega^\mathrm{R,A})^{2}-\varepsilon_{0\mathrm{s}}^{2}}L_{0}^{0}(\frac{|\mathbf{r}_{\perp}-\mathbf{r}_{\perp}^{\prime}|^{2}}{2l_\mathrm{B}^{2}})P_{-\mathrm{s}}\\
& + \sum_{n\ge1}\frac{1}{(\omega^\mathrm{R,A})^{2} - \varepsilon_{n}^{2}}[(\omega^{R,A}+\mathcal{M}_\mathrm{s}\sigma_{z})(L_{n-\frac{1+\mathrm{s}}{2}}^{0}(\frac{|\mathbf{r}_{\perp}-\mathbf{r}_{\perp}^{\prime}|^{2}}{2l_\mathrm{B}^{2}})P_{+}+L_{n-\frac{1-\mathrm{s}}{2}}^{0}(\frac{|\mathbf{r}_{\perp}-\mathbf{r}_{\perp}^{\prime}|^{2}}{2l_\mathrm{B}^{2}})P_{-})\\
& + \sqrt{n}\eta\sqrt{\frac{2}{n}}
\left[\begin{array}{cc}
0 & \frac{i(x-x^{\prime})-(y-y^{\prime})}{2l_\mathrm{B}}\\
\frac{i(x-x^{\prime})+(y-y^{\prime})}{2l_\mathrm{B}} & 0
\end{array}
\right]
e^{-\frac{|\mathbf{r}_{\perp}-\mathbf{r}_{\perp}^{\prime}|^{2}}{4l_\mathrm{B}^{2}}}L_{n-1}^{1}(\frac{|\mathbf{r}_{\perp}-\mathbf{r}_{\perp}^{\prime}|^{2}}{2l_\mathrm{B}^{2}})]\Big\}.
\end{align*}
with $L_{n}^{\alpha}(x)$ as the generalized Laguerre polynomials.
The Fourier transform of translational invariant part $\widetilde{G}(\omega;\mathbf{r}-\mathbf{r}^{\prime})$
can be obtained as 
\begin{equation}
\widetilde{G}_\mathrm{s}^\mathrm{R,A}(\omega,\mathbf{k})=\int d\mathbf{r}e^{-i\mathbf{k}\cdot(\mathbf{r}-\mathbf{r}^{\prime})}\widetilde{G}_\mathrm{s}^\mathrm{R,A}(\omega;\mathbf{r}-\mathbf{r}^{\prime})=e^{-\mathbf{k}_{\perp}^{2}l_\mathrm{B}^{2}}\sum_{n=0}^{\infty}(-1)^{n}\frac{S_\mathrm{ns}^\mathrm{R,A}(\omega,\mathbf{k})}{(\omega^\mathrm{R,A})^{2}-\varepsilon_\mathrm{ns}^{2}}\label{eq:Green's_function}
\end{equation}
with 
\begin{equation}
S_\mathrm{ns}^\mathrm{R,A}(\omega,\mathbf{k})=2(\omega^\mathrm{R,A}+\mathcal{M}_\mathrm{s}\sigma_\mathrm{z})[L_\mathrm{n}(x)P_{-\mathrm{s}}-L_{\mathrm{n}-1}(x)P_\mathrm{s}]-4\mathbf{k}_{\perp}\cdot\boldsymbol{\sigma}L_{\mathrm{n}-1}^{1}(x)\label{eq:S_expression}
\end{equation}
where $x=2|\mathbf{k}_{\perp}|^{2}l_{B}^{2}$ with $\mathbf{k}_{\perp}=(k_\mathrm{x},k_\mathrm{y})$
and the projection operator is $P_\mathrm{s}=(1+s\sigma_\mathrm{z})/2$.

\paragraph*{Kubo Formula for Dirac Materials in a Finite Magnetic Field}

The frequency-dependent electrical conductivity tensor is calculated
using the Kubo formula: $\sigma_\mathrm{ab}(\Omega)=\frac{\mathrm{Im}\Pi_\mathrm{ab}^{R}(\Omega+i0)}{\Omega}$
where $\Pi_\mathrm{ab}^\mathrm{R}$ is the retarded current-current correlation
function obtained by analytically continuing the imaginary time expression:
\[
\Pi_\mathrm{ab}(i\Omega_\mathrm{m})=\frac{1}{V}\int_{0}^{\beta}d\tau e^{i\Omega_\mathrm{m}\tau}\langle T_{\tau}J_\mathrm{a}(\tau)J_\mathrm{b}(0)\rangle=\frac{1}{\beta V}\sum_{\mathbf{k},i\omega_\mathrm{n},\mathrm{s}}\mathrm{Tr}[\widetilde{G}_\mathrm{s}(i\omega_\mathrm{n},\mathbf{k})\hat{V}_\mathrm{a}^\mathrm{s}\widetilde{G}_\mathrm{s}(i\omega_\mathrm{n}+i\Omega_\mathrm{m},\mathbf{k})\hat{V}_\mathrm{b}^{s}]
\]
where $\hat{V}_\mathrm{a}^\mathrm{s}=\frac{1}{\hbar}\partial H_\mathrm{s}/\partial k_\mathrm{a}$ is
the current operator for subblock $s$. By substituting Eq. (\ref{eq:Green's_function})
and performing the analytical continuation $i\Omega_\mathrm{m}\to\Omega+i0$,
the magnetoconductivity can be evaluated as\textsuperscript{\citep{gusynin2006transport}}

\begin{align*}
\sigma_\mathrm{ab}(\Omega) & =\frac{e^{2}v^{2}}{2\pi\Omega}\mathrm{Re}\int_{-\infty}^{\infty}d\omega\int\frac{d^{3}\mathbf{k}}{(2\pi)^{3}}e^{-2\mathbf{k}_{\perp}^{2}l_\mathrm{B}^{2}}\sum_{\mathrm{s}=\pm}s\sum_{n,m=0}^{\infty}(-1)^\mathrm{n+m}tr\{[f_\mathrm{F}(\omega)-f_\mathrm{F}(\omega^{\prime})]\\
 & \times\frac{\sigma_\mathrm{a}S_\mathrm{ns}^\mathrm{R}(\omega^{\prime},\mathbf{k})\sigma_\mathrm{b}S_\mathrm{ms}^\mathrm{A}(\omega,\mathbf{k})}{\{[\omega^{\prime \mathrm{R}}]^{2}-\varepsilon_\mathrm{ns}^{2}\}\{[\omega^\mathrm{A}]^{2}-\varepsilon_\mathrm{ms}^{2}\}}-f_\mathrm{F}(\omega)\frac{\sigma_\mathrm{a}S_\mathrm{ns}^\mathrm{R}(\omega^{\prime},\mathbf{k})\sigma_\mathrm{b}S_\mathrm{ms}^\mathrm{R}(\omega,\mathbf{k})}{\{[\omega^{\prime \mathrm{R}}]^{2}-(\varepsilon_\mathrm{ns})^{2}\}\{[\omega^\mathrm{R}]^{2}-(\varepsilon_\mathrm{ms})^{2}\}}\\
 & +f_\mathrm{F}(\omega^{\prime})\frac{\sigma_\mathrm{a}S_\mathrm{ns}^\mathrm{A}(\omega^{\prime},\mathbf{k})\sigma_\mathrm{b}S_\mathrm{ms}^\mathrm{A}(\omega,\mathbf{k})}{\{[\omega^{\prime \mathrm{A}}]^{2}-(\varepsilon_\mathrm{ns})^{2}\}\{[\omega^\mathrm{A}]^{2}-(\varepsilon_\mathrm{ms})^{2}\}}\}
\end{align*}
where $f_\mathrm{F}$ represents the Fermi-Dirac distribution, $\omega^{\prime}=\omega+\Omega$,
and $\omega^\mathrm{R,A}=\omega\pm i\Gamma$.By performing the trace while
integrating out $\mathbf{k}_{\perp}$, we can derive a more manageable
form of $\sigma_{ab}$.

\paragraph*{Small magnetic field expansion of the Kubo-Streda formula}

In the Kubo-Streda formalism, small field corrections to conductivity
can be derived from the expansion of the Green's function in Eq. (\ref{eq:Green's_function}).
To linear order in the background electromagnetic fields, the Green's
function takes the form: 
\[
G(\mathbf{r},\mathbf{r}^{\prime})=G_{0}(\mathbf{r}-\mathbf{r}^{\prime})+G_{1}(\mathbf{r},\mathbf{r}^{\prime})+....
\]
 Due to translational invariance in the absence of background fields,
the zeroth-order Green's function depends solely on the difference
$\mathbf{r}-\mathbf{r}^{\prime}$, which is not true for the first-order
part of the Green\textquoteright s function. The Fourier transform
of $G_{0}$ can be derived from the model Hamiltonian: $G_{0}^\mathrm{R/A}(\mu,\mathbf{k})=(\mu\pm i\Gamma-H_{0})^{-1}$.
The first-order correction to the Green's function arising from the
magnetic field can be expressed as\textsuperscript{\citep{gorbar2017origin}}: 
\[
G_{1}(\mathbf{r},\mathbf{r}^{\prime})=\int d\mathbf{r}^{\prime\prime}G_{0}(\mathbf{r}-\mathbf{r}^{\prime\prime})H_\mathrm{int}(\mathbf{r}^{\prime\prime})G_{0}(\mathbf{r}^{\prime\prime}-\mathbf{r}^{\prime}).
\]
where the interaction Hamiltonian $H_\mathrm{int}=\mathbf{J}\cdot\mathbf{A}$
accounts for the orbital effect of the magnetic field. The spin effect
of the magnetic field is fully incorporated in $H_{0}$ without any
approximations. Here, we are considering only the translationally
invariant part of the Green's function, which can be obtained by 
\[
\widetilde{G}_{1}=-i\epsilon_\mathrm{cde}\frac{B_{e}}{2}G_{0}V_\mathrm{c}G_{0}V_\mathrm{d}G_{0}
\]
Thus, the first-order correction to the conductivity due to the orbital
effect of the magnetic field is given by:
\begin{align*}
\sigma_\mathrm{ab}^{(1)} & =\frac{e^{2}v^{2}}{\pi}\int[d\mathbf{p}]\mathrm{Re}tr[V_\mathrm{a}G_{0}^\mathrm{R}V_\mathrm{b}\widetilde{G}_{1}^\mathrm{A}]\\
 & =\frac{e^{2}v^{2}}{\pi}\epsilon_\mathrm{cde}\frac{B_\mathrm{e}}{2}\int[d\mathbf{p}]\mathrm{Im}tr[V_\mathrm{a}G_{0}^{R}V_\mathrm{b}G_{0}^\mathrm{A}V_{c}G_{0}^\mathrm{A}V_\mathrm{d}G_{0}^\mathrm{A}]
\end{align*}
with $\int[d\mathbf{p}]=\int\frac{d^{3}\mathbf{p}}{(2\pi)^{3}}$.
Here, $a,b,c,d,...$ denote the Cartesian components, and we adopt
the Einstein summation convention for repeated indices. The trace
can be evaluated by expressing the Green's function in terms of the
eigenenergy basis $G_{0}^\mathrm{R/A}(\mathbf{p})=\sum_{\alpha}|u_{\alpha\mathbf{p}}\rangle G_{0,\alpha}^\mathrm{R/A}(\mathbf{p})\langle u_{\alpha\mathbf{p}}|$
where $H_{0}|u_{\alpha\mathbf{p}}\rangle=\varepsilon_{\alpha\mathbf{p}}|u_{\alpha\mathbf{p}}\rangle$
and $G_{0,\alpha}^\mathrm{R/A}(\mathbf{p})=(\mu\pm i\Gamma-\varepsilon_{\alpha\mathbf{p}})^{-1}$.
In the analysis of a two-band model, the first-order correction to
the conductivity tensor, $\sigma_\mathrm{ab}^{(1)}$, can be divided into
two distinct contributions:

1. Two of the band indices of the Green's functions are identical
$\sigma_\mathrm{ab}^{(1,\mathrm{i})}$: 

\begin{align}
\sigma_{ab}^{(1,\mathrm{i})} & =\frac{e^{2}B_\mathrm{e}}{2\pi}\epsilon_\mathrm{cde}\int[d\mathbf{p}]\sum_{\beta\ne\alpha}\mathrm{Im}[G_{0\alpha}^\mathrm{R}G_{0\alpha}^\mathrm{A}(G_{0\beta}^\mathrm{A})^{2}\nonumber \\
 & \times(V_\mathrm{a}^{\beta\alpha}V_\mathrm{b}^{\alpha\alpha}V_\mathrm{c}^{\alpha\beta}V_\mathrm{d}^{\beta\beta}+V_\mathrm{a}^{\beta\alpha}V_\mathrm{b}^{\alpha\beta}V_\mathrm{c}^{\beta\alpha}V_\mathrm{d}^{\alpha\beta}+V_\mathrm{a}^{\alpha\alpha}V_\mathrm{b}^{\alpha\beta}V_\mathrm{c}^{\beta\beta}V_\mathrm{d}^{\beta\alpha})].\label{eq:sigma_ii}
\end{align}

2. Three of the band indices of the Green's functions are identical
$\sigma_{ab}^{(1,\mathrm{ii})}$ :
\begin{align}
\sigma_\mathrm{ab}^{(1,\mathrm{ii})} & =\frac{e^{2}B_\mathrm{e}}{2\pi}\epsilon_\mathrm{cde}\int[d\mathbf{p}]\sum_{\beta\ne\alpha}\mathrm{Im}[G_{0\alpha}^\mathrm{R}(G_{0\alpha}^\mathrm{A})^{2}G_{0\beta}^\mathrm{A}\nonumber \\
 & \times(V_\mathrm{a}^{\beta\alpha}V_\mathrm{b}^{\alpha\alpha}V_{c}^{\alpha\alpha}V_\mathrm{d}^{\alpha\beta}+V_\mathrm{a}^{\alpha\alpha}V_\mathrm{b}^{\alpha\alpha}V_\mathrm{c}^{\alpha\beta}V_\mathrm{d}^{\beta\alpha}+V_\mathrm{a}^{\alpha\alpha}V_\mathrm{b}^{\alpha\beta}V_\mathrm{c}^{\beta\alpha}V_\mathrm{d}^{\alpha\alpha})].\label{eq:sigma_iii}
\end{align}
In the weak scattering limit ($\Gamma\to0$), the product of the four
Green's functions provides the leading contribution at $\mu\sim\varepsilon_{\alpha}$
or $\varepsilon_{\beta}$, and can be approximated using Dirac delta
functions:
\begin{align}
G_{0\alpha}^\mathrm{R}G_{0\alpha}^\mathrm{A}(G_{0\beta}^\mathrm{A})^{2}\simeq i\Big[\frac{2\pi}{\varepsilon_{\alpha\beta}^{3}}\delta(\mu-\varepsilon_{\alpha})-\frac{\pi}{(2\mu-\varepsilon_{\alpha}-\varepsilon_{\beta})\varepsilon_{\alpha\beta}}\delta^{\prime}(\mu-\varepsilon_{\beta})\Big]+ & \frac{\pi}{\varepsilon_{\alpha\beta}^{2}\Gamma}\delta(\mu-\varepsilon_{\alpha})\label{eq:RAAA}
\end{align}
and
\begin{align}
G_{0\alpha}^\mathrm{R}(G_{0\alpha}^\mathrm{A})^{2}G_{0\beta}^\mathrm{A}\simeq & i\Big[\frac{\pi}{2\Gamma^{2}\varepsilon_{\alpha\beta}}\delta(\mu-\varepsilon_{\alpha})-\frac{\pi}{\varepsilon_{\alpha\beta}^{3}}[\delta(\mu-\varepsilon_{\alpha})+\delta(\mu-\varepsilon_{\beta})]+\frac{\pi}{4\varepsilon_{\alpha\beta}}\delta^{\prime\prime}(\varepsilon_{\alpha}-\mu)\Big]\nonumber \\
 & +\frac{\pi}{2\varepsilon_{\alpha\beta}\Gamma}\delta^{\prime}(\mu-\varepsilon_{\alpha})-\frac{\pi}{\varepsilon_{\alpha\beta}^{2}\Gamma}\delta(\mu-\varepsilon_{\alpha})\label{eq:RAAA2}
\end{align}
where $\delta^{\prime}(\mu-\varepsilon)=\frac{\partial\delta(\mu-\varepsilon)}{\partial\varepsilon}$
, $\delta^{\prime\prime}(\mu-\varepsilon)=\frac{\partial^{2}\delta(\mu-\varepsilon)}{\partial\varepsilon^{2}}$
and $\varepsilon_{\alpha\beta}\equiv\varepsilon_{\alpha}-\varepsilon_{\beta}$. 

By substituting Eqs. (\ref{eq:RAAA}) and (\ref{eq:RAAA2}) into Eqs.
(\ref{eq:sigma_ii}) and (\ref{eq:sigma_iii}), we can obtain the
current $j_\mathrm{a}=\chi_\mathrm{abc}E_\mathrm{b}B_\mathrm{c}$ in the order $O(EB)$, where
the response tensor is given by $\chi_\mathrm{abc}=\chi_\mathrm{abc}^{(2)}+\chi_\mathrm{abc}^{(1)}+\chi_\mathrm{abc}^{(0)}$,
which are proportional to $\propto\tau^{2},\tau^{1},\tau^{0}$ respectively.
The magnetoconductivity for different orders of relaxation time can
be expressed as $\sigma_\mathrm{ab}^{(\mathrm{i})}=\chi_\mathrm{abc}^{(\mathrm{i})}B_\mathrm{c}$ with $i=2,1,0$. 

The explicit expressions for magnetoconductivity $\sigma_\mathrm{ab}^{(\mathrm{i})}$
are given by:

\begin{align}
\sigma_\mathrm{ab}^{(2)} & \simeq-\frac{e^{2}}{4\Gamma^{2}\hbar}\int[d\mathbf{p}]\frac{\partial f}{\partial\epsilon_{0}}[n_\mathrm{ac}V_\mathrm{b}(\mathbf{B}\times\mathbf{V})_\mathrm{c}-(a\longleftrightarrow b)],\label{eq:tau2}
\end{align}

\begin{align}
\sigma_\mathrm{ab}^{(1)} & \simeq-\frac{e^{2}}{\Gamma\hbar}\int[d\mathbf{p}]\{-\frac{\partial f}{\partial\varepsilon_{0}}\frac{1}{8}\mathrm{Re}\langle\{\hat{V}_\mathrm{a},\hat{V}_\mathrm{b}\}\rangle(\boldsymbol{\Omega}\cdot\mathbf{B})-\frac{1}{4}\frac{\partial^{2}f}{\partial\varepsilon_{0}^{2}}B_\mathrm{a}V_\mathrm{b}(\mathbf{m}\cdot\mathbf{V})+(a\longleftrightarrow b)\},\label{eq:tau1}
\end{align}
\begin{align}
{\normalcolor \sigma_\mathrm{ab}^{(0)}} & \mathrel{\normalcolor \simeq}{\normalcolor -\frac{e^{2}}{\hbar}\int[d\mathbf{p}]\{\frac{\partial f}{\partial\varepsilon_{0}}[{\color{red}{\normalcolor B_\mathrm{c}F_\mathrm{cb}V_\mathrm{a}+\frac{1}{2}\epsilon_\mathrm{abc}\Omega_\mathrm{c}(\mathbf{m}\cdot\mathbf{B}}\mathclose{\normalcolor )}}}{\normalcolor \mathbin{\color{black}-}{\color{black}{\color{blue}\mathinner{\color{black}\frac{1}{2}}{\color{black}g_\mathrm{ac}(\mathbf{B}\times\nabla_{\mathbf{p}}V_\mathrm{b}}\mathclose{\color{black})_\mathrm{c}}}}}\mathclose{\normalcolor ]}\nonumber\\
{\normalcolor } & \mathbin{\normalcolor +}{\normalcolor \frac{\partial^{2}f}{\partial\varepsilon_{0}^{2}}[{\color{blue}\mathinner{\color{black}\frac{\mathrm{Im}\langle[\hat{V}_\mathrm{a},\hat{V}_\mathrm{b}]\rangle(\mathbf{m}\cdot\mathbf{B})}{8\mu}}}}\mathclose{\normalcolor ]}\mathbin{\normalcolor +}{\normalcolor \frac{\partial^{3}f}{\partial\varepsilon_{0}^{3}}\frac{1}{8}n_\mathrm{ac}V_\mathrm{b}(\mathbf{B}\times\mathbf{V})_{c}-(a\longleftrightarrow b)\}.}\label{eq:tau0}
\end{align}
Here, $\epsilon_\mathrm{abc}$ represents the Levi-Civita symbol. For a particular
band with index $\beta,$ the anomalous orbital polarizability (AOP)
is defined as $F_\mathrm{ba}=2\mathrm{Re}\frac{\mathcal{M}_\mathrm{b}^{\beta\alpha}\mathcal{A}_\mathrm{a}^{\alpha\beta}}{\varepsilon_{\beta\alpha}}+\frac{1}{2}\epsilon_\mathrm{bcd}\partial_{p_\mathrm{c}}g_\mathrm{ad}$
where $\mathcal{A}_\mathrm{a}^{\alpha\beta}=\langle u_{\alpha\mathbf{p}}|i\partial_{p_\mathrm{a}}|u_{\beta\mathbf{p}}\rangle$
is the unperturbated interband Berry connection, $\mathcal{M}_\mathrm{a}^{\alpha\beta}=\frac{1}{2}\epsilon_{abc}(V_\mathrm{b}^{\alpha\alpha}+V_\mathrm{b}^{\beta\beta})\mathcal{A}_\mathrm{c}^{\alpha\beta}$
is the interband orbital magnetic moments, with $V_\mathrm{b}^{\alpha\beta}$
being the matrix elements of velocity operator, $g_\mathrm{ab}=\mathrm{Re}(\mathcal{A}_\mathrm{a}^{\alpha\beta}\mathcal{A}_\mathrm{b}^{\beta\alpha})$
as the quantum metric tensor, and $\varepsilon_\mathrm{\beta\alpha}=\varepsilon_{\beta}-\varepsilon_{\alpha}.$
The intraband orbital magnetic moment $m_\mathrm{c}=-\frac{1}{2}\epsilon_\mathrm{abc}\mathrm{Im}\langle\partial_{p_\mathrm{a}}u_{\beta\mathbf{p}}|(\varepsilon_{\beta}-\hat{H}_{0})|\partial_{p_\mathrm{b}}u_{\beta\mathbf{p}}\rangle$
and the real part of the quantity $n_\mathrm{ab}=\mathrm{Re}\langle\partial_{p_\mathrm{a}}u_{\beta\mathbf{p}}|(\varepsilon_{\beta}-\hat{H}_{0})|\partial_{p_\mathrm{b}}u_{\beta\mathbf{p}}\rangle$. 

\backmatter

\bmhead{Data availability statement}

All data generated or analysed during this study are included in this published article (and its supplementary information files).

\bmhead{Acknowledgments}

This work was supported by the National Natural Science Foundation
of China (Grants No. 12304192, No. 11904062, and No. 12374175), the
National Key R\&D Program of China (Grant No. 2019YFA0308603), the
Guangdong Basic and Applied Basic Research Foundation (Grants No.
2024A1515010430, No. 2024A1515012689, and No. 2023A1515140008), Guangdong Province Introduced Innovative R\&D Team Program (Grant No. 2023QN10X136), the
Research Grants Council, University Grants Committee, Hong Kong (Grants
No. C7012-21G and No. 17301823), Quantum Science Center of Guangdong-Hong
Kong-Macao Greater Bay Area (Grant No. GDZX2301005). H.W.W was also
supported by the Sichuan Science and Technology Program (Grant No.
2024NSFSC1376), the China Postdoctoral Science Foundation (Grant No.
2023M740525), and the International Postdoctoral Exchange Fellowship
Program (Grant No. YJ20220059).

\bmhead{Author contributions}

H.W. W.,W. S., and  S.-Q. S conceived the project. B. F. and H.X.  performed the theoretical analysis and simulation. B. F. and H.X.  wrote the manuscript with inputs from all authors. All authors contributed to the discussion of the results. 

\bmhead{Competing interests}

The authors declare no competing interests.

\bmhead{Additional information}

$\mathbf{Supplementary}$ 
$\mathbf{information}$ The online version contains
supplementary material available at XXX.

$\mathbf{Correspondence}$ $\mathbf{and}$ $\mathbf{requests}$ $\mathbf{for}$ $\mathbf{materials}$ should be addressed to Huan-Wen Wang, Wenyu Shan, or Bo Fu.

\begin{appendices}




\end{appendices}



\bibliography{refer}



\end{document}